\documentclass[%
 aip,
 amsmath,amssymb,
 reprint,%
]{revtex4-1}

\usepackage{graphicx}
\usepackage{dcolumn}
\usepackage{bm}
\usepackage{textcomp}

\usepackage[utf8]{inputenc}
\usepackage[T1]{fontenc}
\usepackage{mathptmx}
\usepackage{braket}

\usepackage{ragged2e}
\usepackage[format=plain,
      justification=RaggedRight,
      singlelinecheck=false]
     {caption}

\usepackage{subcaption}
\usepackage{floatrow}
\begin{document}

\preprint{AIP/123-QED}

\title[Red- and blue-detuned magneto-optical trapping with liquid crystal variable retarders]{Red- and blue-detuned magneto-optical trapping\\with liquid crystal variable retarders}

\author{B. Piest}
 \email{piest@iqo.uni-hannover.de (corresponding author)}
 \affiliation{Institut für Quantenoptik, Gottfried Wilhelm Leibniz Universität, Welfengarten 1, 30167 Hannover, Germany.}%

\author{V. Vollenkemper}%
 \affiliation{Institut für Quantenoptik, Gottfried Wilhelm Leibniz Universität, Welfengarten 1, 30167 Hannover, Germany.}%
\author{J. Böhm}
\affiliation{Institut für Quantenoptik, Gottfried Wilhelm Leibniz Universität, Welfengarten 1, 30167 Hannover, Germany.}%
\author{A. Herbst}
\affiliation{Institut für Quantenoptik, Gottfried Wilhelm Leibniz Universität, Welfengarten 1, 30167 Hannover, Germany.}%
\author{E. M. Rasel}
\affiliation{Institut für Quantenoptik, Gottfried Wilhelm Leibniz Universität, Welfengarten 1, 30167 Hannover, Germany.}%

\date{\today}

\begin{abstract}
We exploit red- and blue-detuned magneto optical trapping (MOT) of $^{87}$Rb benefitting from a simplified setup based on liquid crystal variable retarders (LCVR). To maintain the trapping forces when switching from a red- to a blue-detuned MOT, the circularity of the cooling beams needs to be reversed. LCVRs allow fast polarization control and represent compact, simple and cost-efficient components which can easily be implemented in existing laser systems. This way, we achieve a blue-detuned type-II MOT for $^{87}$Rb atoms with sub-Doppler temperatures of 44.5$\,$\textmu K. The phase space density is increased by more than two orders of magnitude compared to the standard red-detuned type-I MOT. The setup can readily be transferred to any other system working with $^{87}$Rb.
\end{abstract}

\maketitle

\section{\label{sec:Introduction}Introduction}
\begin{figure}
\includegraphics[scale=0.5]{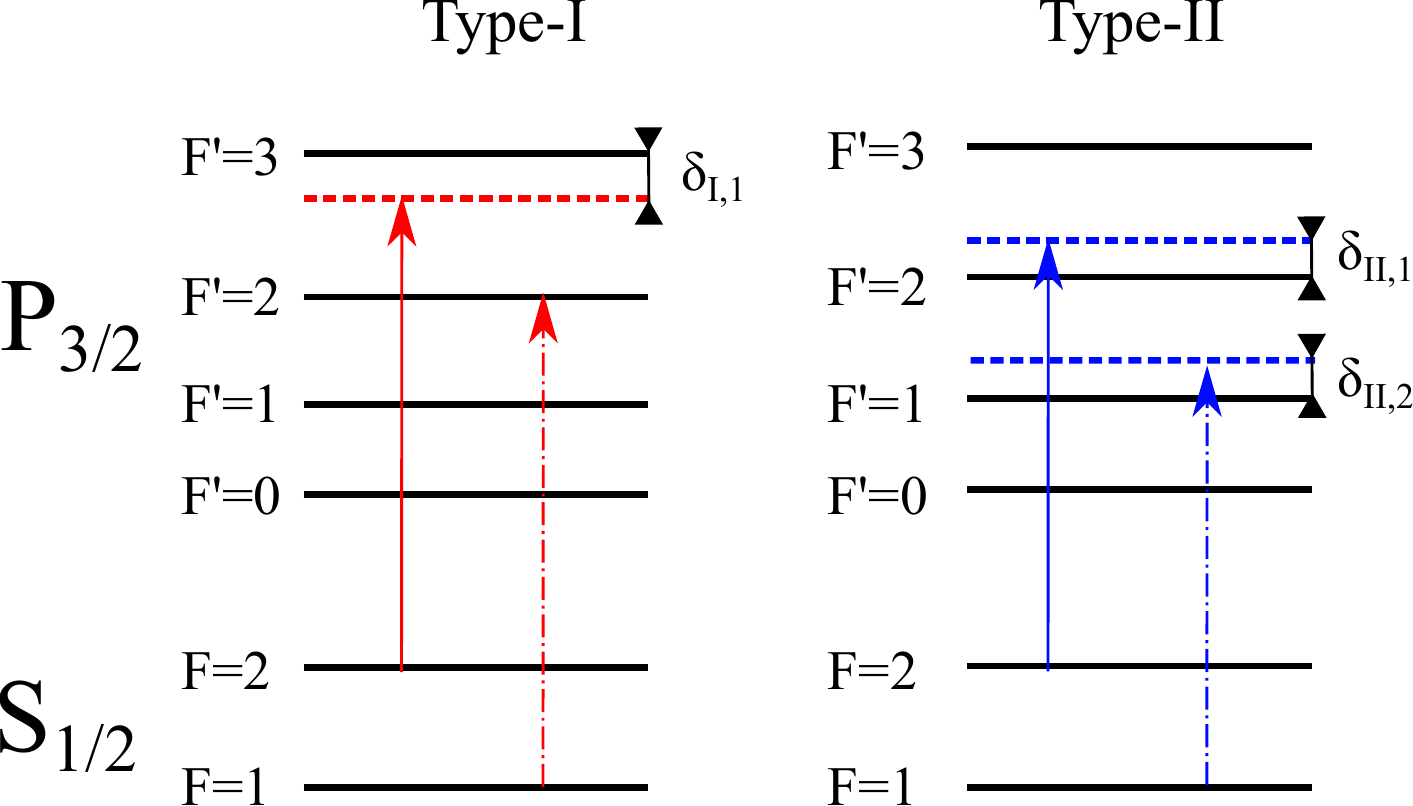}
\caption{Frequency detunings used for the type-I and type-II MOT, respectively. The experimentally determined ideal frequency detunings are $\delta_{\mathrm{I},1}=-11\,$MHz, $\delta_{\mathrm{II},1}=46\,$MHz and $\delta_{\mathrm{II},2}=22\,$MHz. ECDL-1 (ECDL-2) are indicated by solid (dashed) arrows.}
\label{fig:LCVRs}
\end{figure}

Trapped cold atoms with high phase space densities (PSD) are widely used in a wealth of experiments in atomic, optical and molecular physics, cold chemistry and nano physics. Typical applications range from inertial sensing \cite{Geiger2020}, optical frequency standards \cite{Ludlow2015} to ultrasensitive isotope trace analysis \cite{Lu2010} or bright sources for ion beams \cite{Steele2017}. The most commonly used approach to generate cold atoms is the use of a so-called type-I MOT \cite{Phillips1998} although the achievable PSD in conventional MOTs is constrained by atom collisions \cite{Prentiss1988}, photon reabsorption \cite{Walker1990} and the lack of sub-Doppler cooling forces in magnetic fields\cite{Walhout1992}. Among various approaches\cite{Ketterle1993, Radwell2013}, blue-detuned type-II MOTs circumvent these limitations by enabling sub-Doppler forces even in non-zero magnetic fields.\\
MOTs can be realized by using either type-I or type-II transitions which refer to the involved atomic substates of the idealized two-level system. In type-I MOTs, the total angular momentum of the excited state $F'$ is larger compared to that of the ground state $F$, thus satisfying $F'=F+1$. Type-II MOTs in contrast make use of transitions where $F'=F$ or $F'=F-1$ (see fig. \ref{fig:LCVRs} for $^{87}$Rb). When using blue-detuned light fields in a type-II MOT, the sub-Doppler cooling forces are substantially enhanced albeit capturing and cooling atoms with higher temperatures is not feasible \cite{Devlin2016}.
Thus, to efficiently access the sub-Doppler cooling range, a pre-cooled atom ensemble loaded via a standard type-I MOT is required \cite{Jarvis2018}. Furthermore, the polarization of the light fields or the magnetic field gradients have to be reversed when switching between the two MOT types to maintain the position depend trapping forces which requires sophisticated optical setups \cite{Devlin2016,Jarvis2018}.\\
Here, we demonstrate a low expenditure method based on switching a conventional red-detuned MOT to a blue-detuned type-II MOT using LCVRs. With this marginally altered setup we increase the PSD of a $^{87}$Rb type-I MOT by more than two orders of magnitude. We also show the feasibility of cooling and trapping of $^{41}$K or $^{39}$K in a type-II MOT, although no improvement compared to type-I MOTs is observed.\\

Due to its simplicity, the setup is especially suitable for compact devices such as mobile atom sensors \cite{Hinton2017,Bidel2018} or space-borne experiments \cite{Rudolph2015,Elsen2021,Frye2021} and can easily and cost-efficiently be implemented in existing setups.

\captionsetup[sub]{font=normalsize,labelfont={bf,sf}}
\begin{figure*}[t]
\begin{subfigure}{.5\textwidth}
    \caption{}
    \includegraphics[scale = 0.68]{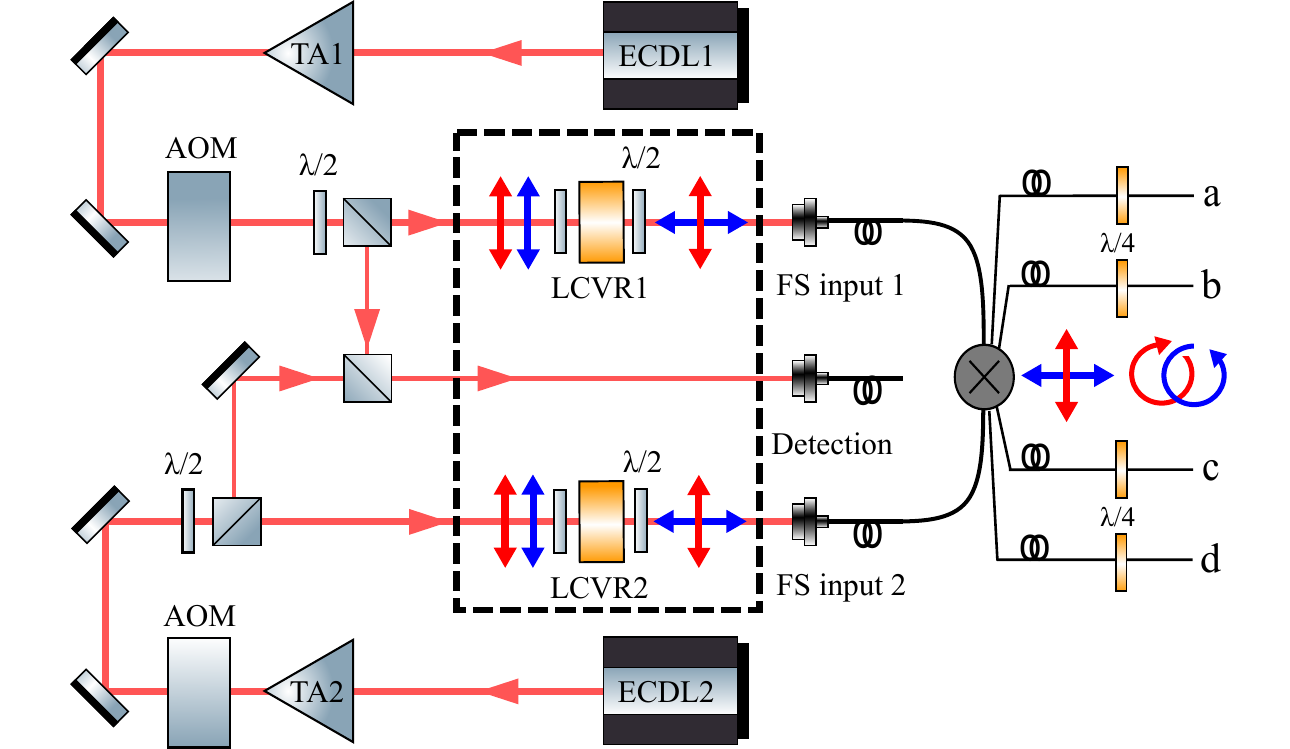}
\end{subfigure}%
\begin{subfigure}{.39\textwidth}
    \caption{}
    \includegraphics[scale=0.15]{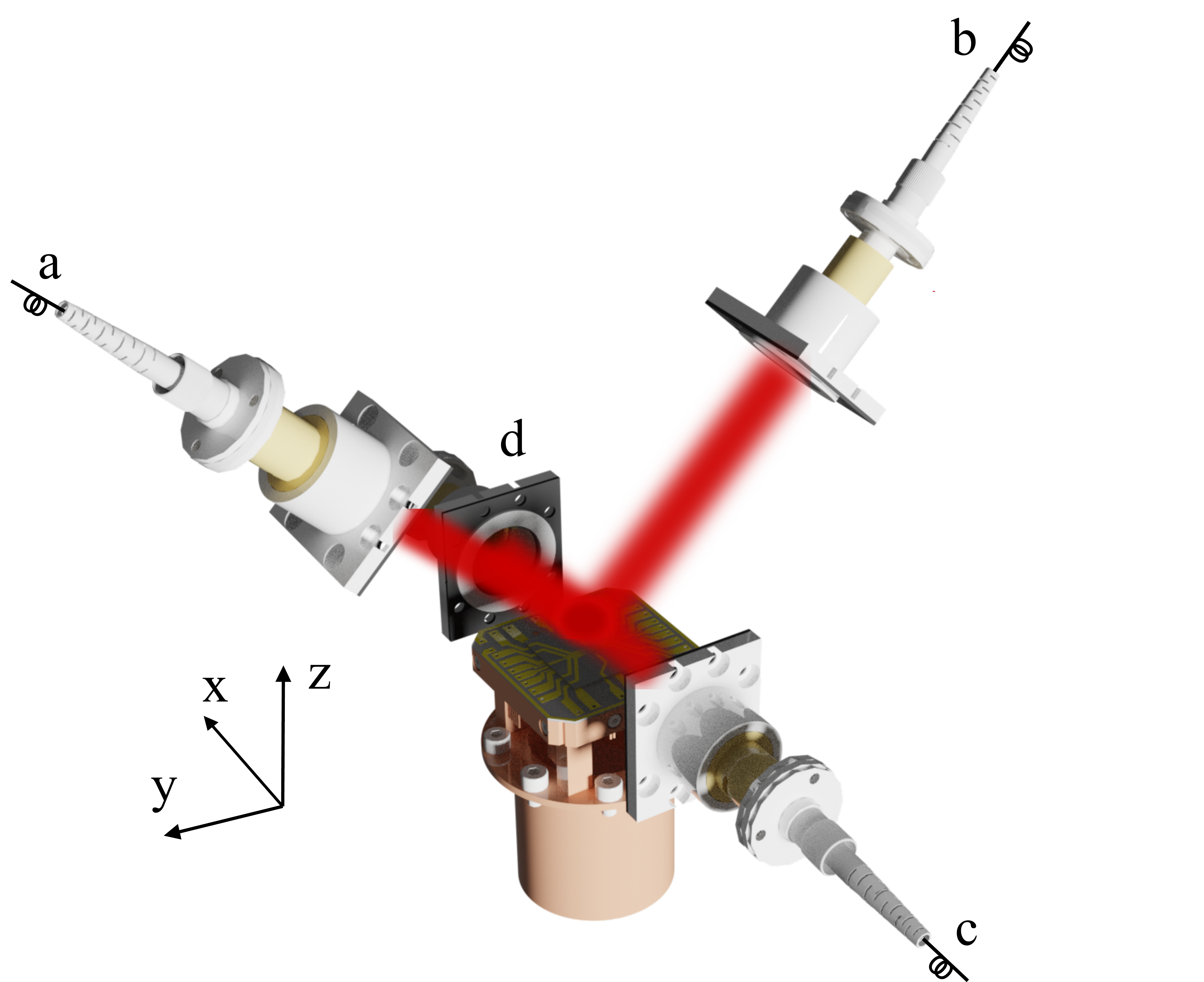}
\end{subfigure}{}
\caption{(a) Setup of laser module and fiber splitter with respective fiber outputs a-d. TA: Tapered amplifier, FS: fiber splitter, AOM: acousto-optical modulator. The red (blue) arrows indicate the direction of polarization for the red- (blue-) detuned MOT. (b) Setup of the experimental chamber with the atom chip and external beam collimators. Shown are the four MOT laser beams and the atom chip. The beams a and b are reflected on the chip surface which forms a six-beam MOT configuration.}
\label{fig:setup}

\end{figure*}

\section{Setup}
The experimental setup shown in fig. \ref{fig:setup} has initially been designed to operate only type-I $^{87}$Rb MOTs. The experimental chamber features an atom chip housed in a vacuum system and external Helmholtz coils to generate the required magnetic fields in quadrupole configuration. Four collimators are fixed around the vacuum chamber to form the MOT light fields, two of them being reflected on the atom chip surface \cite{Folman2000, Wildermuth2004}. The details of the system are described in \cite{Elsen2021}. The laser light is delivered via polarization maintaining (PM) optical fibers to the respective collimators where it passes a quarter-waveplate to provide circular polarization. It is generated by two home-built narrow linewidth external cavity diode lasers (ECDL 1 and ECDL 2 in fig. \ref{fig:setup} (a)) which operate initially at the cooling ($\ket{F=2}\rightarrow \ket{F'=3}$) and repumping transitions ($\ket{F=1}\rightarrow \ket{F'=2}$) of $^{87}$Rb at 780.242$\,$nm and 780.246$\,$nm, respectively. 
The light is amplified in two single tapered amplifiers with an optical output power of up to 1$\,$W. Subsequently, each beam passes an acousto-optic modulator that is used to control the intensities of the respective light fields. A small fraction is split off to provide light for spatially resolved absorption and fluorescence detection.
The transmitted parts used for atom cooling pass an LCVR (Thorlabs LCC1111-B) and are coupled into a PM fiber-optic splitter. LCVRs consist of a transparent birefringent polymer crystal which molecules align according to an applied AC voltage. Due to the adjustable change of refraction index, they act as waveplates with tunable retardance and fixed orientation.
Here, they are regulated by a square-wave voltage at a frequency of 2$\,$kHz with variable amplitude which allows to continuously rotate the polarization of the incident light. The controller switches between two predefined amplitudes $V1$ and $V2$ by an external trigger. The amplitudes are chosen in order to provide two orthogonal output polarizations indicated by blue (green) arrows in fig. \ref{fig:setup} (a). This allows to make use either of the fast or slow polarization mode of the PM fibers. In combination with the quarter-waveplates after the fiber splitter, this allows to switch between left- and right-circular polarization of the MOT beams.\\
The switching time has been measured in an external setup to be 0.6$\,$ms from $V1$ to $V2$ and 35$\,$ms from $V2$ to $V1$. In principle, this time can be shortened by using additional heating of the LCVRs or by using ultrafast but usually high-cost Pockels cells. Since the detection light is not affected by the LCVRs, only the fast switching direction is used during an experimental sequence. Thus, no further gain from shorter switching times is expected for the here-discussed experiments.
The long-term stability of the retardance is kept below $0.02\lambda$ over more than 100 weeks \cite{datasheet} which obviates any further adjustments of amplitude voltages.\\
This setup allows to reverse the polarization for all beams inside the MOT when switching from the red- to the blue-detuned MOT with only two LCVRs. 
It additionally avoids any mechanical moving parts and is based on cost-efficient components.
\section{Experiments}
\begin{figure*}[t]
\begin{subfigure}{.39\textwidth}
    \caption{}
    \includegraphics[height=4.8cm]{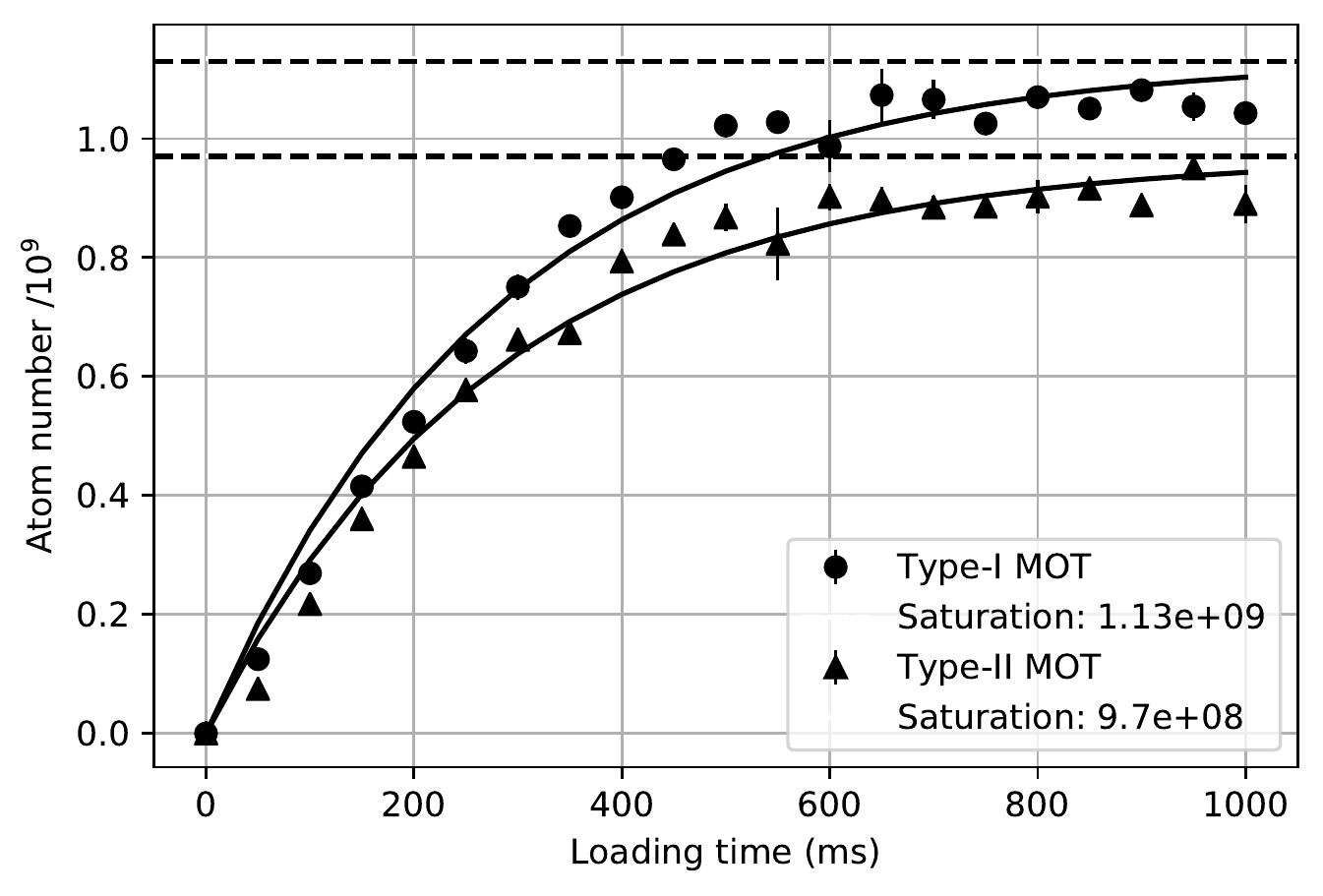}
\end{subfigure}%
\begin{subfigure}{.38\textwidth}
    \caption{}
     \includegraphics[height=4.8cm]{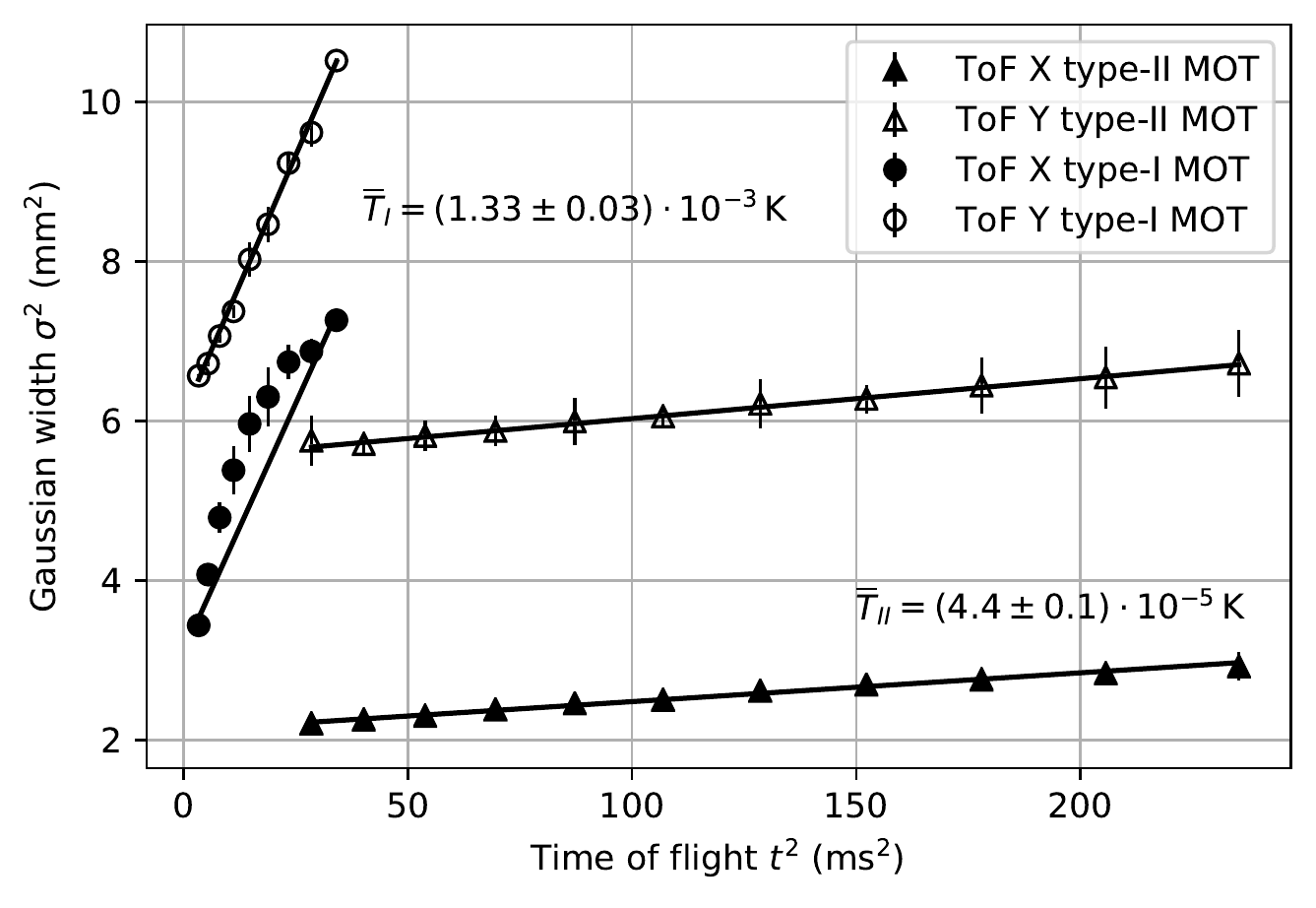}
\end{subfigure}%
\begin{subfigure}{.26\textwidth}
    \caption{}
     \includegraphics[height=4.6cm]{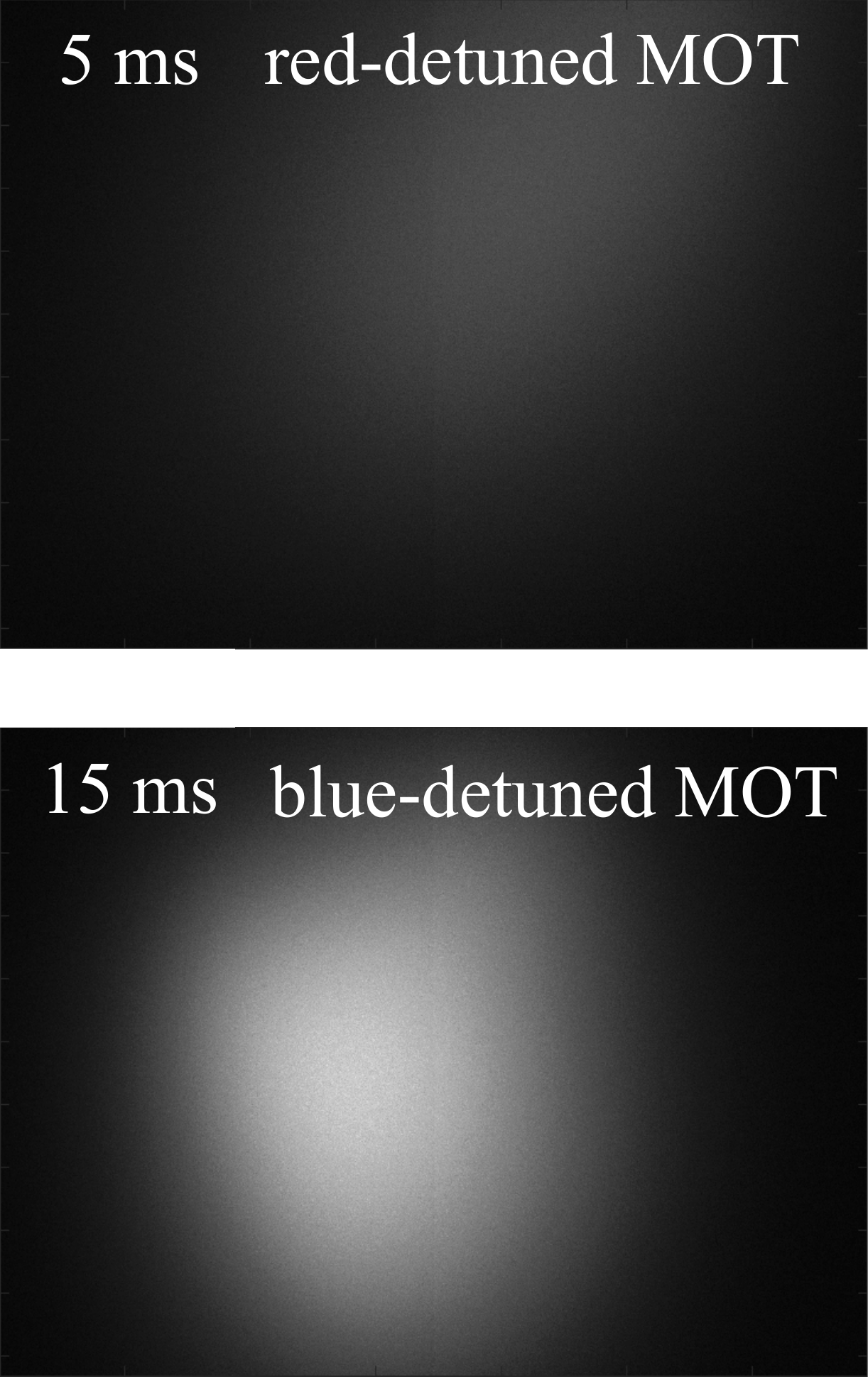}
\end{subfigure}{}
\caption{(a) Comparison of atom number over loading time for both MOTs. Errorbars denote the $1\sigma$ statistical uncertainty over three consecutive experimental runs. Transfer efficiency from type-I MOT to type-II MOT results in $(84.3\pm6.5)\,\%$.  (b) Measurement of ensemble size in dependence of time of flight after release from the respective MOTs in two spatial directions $\sigma_x$ and $\sigma_y$ (see fig. \ref{fig:setup} for coordinate system definition) after a loading time of $500\,$ms. Errorbars denote three standard deviations of three consecutive experimental runs for better visibility. The parameter $\sigma$ is extracted for each direction with a 1D-Gaussian fit function. The expansion is fitted with a linear model (solid lines). The type-II MOT generates lower temperatures than the type-I MOT as revealed by the lower expansion rate. (c) Fluorescence images of a red-detuned (blue-detuned) MOT after free expansions of 5$\,$ms (15$\,$ms). The detected image of the red-detuned MOT exceeds the size of the CCD sensor which deforms its Gaussian envelope along x-direction leading to the observed deviation of the type-I MOT expansion data.} 
\label{fig:experiments}
\end{figure*}
In the following we will discuss our measurements to determine the feasibility of a type-II MOT using the setup presented above and compare our results with the case where we only use a type-I MOT. \\
For the type-I MOT the lasers ECDL 1 and 2 drive the transitions shown in fig. \ref{fig:LCVRs}. The cooling laser is red detuned with $\delta_{I,1}=-11\,$MHz to the type-I cooling transition, the repumper is in resonance with the transition. The cooling and repumping beams operate at peak intensities of $I_{I,1}=18.0\,$mW/cm$^2$ and $I_{I,2}=2.0\,$mW/cm$^2$ for each beam at the position of the atoms, respectively. During the type-I MOT the LCVRs are operated with the amplitude $V1$. After the loading time $T$, the type-I MOT is switched off, followed by a 2$\,$ms long period to stabilize the laser frequencies to the blue detuned cooling transitions shown in fig. \ref{fig:LCVRs}. The acousto-optical modulators are switched off during this period to avoid unintended light forces on the atoms. Also, the LCVRs are turned to $V2$ to reverse the light polarization at the atom position. Subsequently, the lasers ECDL 1 and 2 drive the type-II transitions $\ket{F=2}\rightarrow \ket{F'=2}$ and $\ket{F=1}\rightarrow \ket{F'=1}$ with detunings of $\delta_{\mathrm{II},1}=46\,$MHz and $\delta_{\mathrm{II},2}=22\,$MHz, respectively. Notably the "repumper" transition $\ket{F=1}\rightarrow \ket{F'=1}$ contributes the main part to the light forces \cite{Devlin2016}. Thus the peak intensity of ECDL 2 is increased to its maximum value of $I_{II,2}=8.2\,$mW/cm$^2$ while the peak intensity of ECDL 1 is slightly decreased to $I_{II,1}=13.2\,$mW/cm$^2$ per beam. In the type-II MOT, the atoms are trapped and further cooled for 20$\,$ms. Subsequently, the MOT light and magnetic fields are switched off and the atoms are released into free fall. With a 20$\,$\textmu s light pulse resonant to the transition $\ket{F=1}\rightarrow \ket{F'=2}$, all atoms remaining in $\ket{F=1}$ are pumped into the cycling detection transition $\ket{F=2}\rightarrow \ket{F'=3}$. The atoms are detected via absorption and fluorescence imaging using the fiber output port shown in fig. \ref{fig:setup}(a). Here, absorption imaging is primarily used for atom number detection and fluorescence imaging for expansion measurements in x- and y-direction.

\section{Results}
We characterize the properties of the blue-detuned type-II MOT by determining its transfer efficiency from a type-I MOT and the achievable PSD.
In fig. \ref{fig:experiments}(a), the atom number over the (type-I) MOT loading time $T$ is shown for both MOTs in comparison. The transfer efficiency between the type-I and type-II MOT results in $(84.3\pm6.5)\,\%$ averaged over all loading times. According to simulations of the cooling forces \cite{Jarvis2018}, atoms below a critical velocity of $v\approx 4.5\,$m/s are expected to be trapped in the type-II MOT. 
Given the temperature of the type-I MOT of 1.4$\,$mK, the transfer efficiency should be close to 100\,\% based on a Maxwell-Boltzmann distribution of the atoms in the type-I MOT.
The $1/e$-lifetime of the type-II MOT has been measured to be $(512\pm 17)\,$ms and accounts for a loss of 3.8$\,$\% of the initially trapped atoms during the $20\,$ms type-II MOT period. We attribute the remaining loss fraction to atoms which escape the trapping volume of the type-II MOT during the $2\,$ms switching phase where the ensemble freely expands. Faster frequency jumps could therefore reduce these losses.\\
Fig. \ref{fig:experiments}(b) shows time of flight (ToF) measurements of the Gaussian ensemble width in two spatial directions $\sigma_x$ and $\sigma_y$ of both MOTs for a loading time of $500\,$ms. Here, the MOT light and the quadrupole field are switched off, allowing the ensemble to expand freely until detection. The ensemble size in the xy-plane is detected via fluorescence detection. The expansion of an ensemble containing atoms with mass $m$ is related to its temperature via $T=m\sigma_v^2/k_B$. The parameter $\sigma_v^2$ is given by a least-square fit of the function $\sigma^2(t)=\sigma^2(t=0)+\sigma_v^2 t^2$.
For the red-detuned type-I MOT, a geometric mean temperature of $\overline{T}_\mathrm{I}=(1.33\pm0.03)\,$mK is determined. In the type-II MOT, the ensemble temperature is further reduced to $\overline{T}_{\mathrm{II}}=(44.5\pm0.9)\,$\textmu K which is below the Doppler temperature limit $T_D\approx 140\,$\textmu K \cite{Steck} and indicates efficient sub-Doppler cooling forces. Notably, the ensemble temperature is further reduced for smaller atom densities which is attributed to reduced photon rescattering \cite{Hillenbrand,Townsend}. For example, a temperature of 19.2$\,$\textmu K is reached for a type-II MOT with $2.2\cdot10^8$ atoms.\\
The data obtained from the TOF measurement allows us to calculate the PSD of the ensemble by $\rho = n\cdot\lambda^3_{dB}$. Here, $n$ is the initial peak density of the atom cloud right after release and $$\lambda_{dB}= \sqrt{\frac{2\pi \hbar^2}{m k_B T}}$$ denotes the thermal de-Broglie wavelength. The peak density is evaluated using $n = N/((\sqrt{2\pi})^3 \sigma_x \sigma_y \sigma_z)$ with the total atom number $N$ and the initial Gaussian widths $\sigma_{x,y,z}$ immediately after release. Here, the z-direction is accessible via absorption detection.\\
These equations lead to densities and PSDs of the ensembles after the type-I MOT ($n_{\mathrm{I}}$, $\rho_{\mathrm{I}}$) and after the type-II MOT ($n_{\mathrm{II}}$, $\rho_{\mathrm{II}}$) of

\begin{align*}
    n_{\mathrm{I}} &= (6.8 \pm 0.2) \cdot 10^{9} \, \mathrm{cm^{-3}} \\
    \rho_{\mathrm{I}} &= (9.2 \pm 0.4) \cdot 10^{-10} \\
    n_{\mathrm{II}} &= (1.2 \pm 0.1) \cdot 10^{10} \, \mathrm{cm^{-3}} \\
    \rho_{\mathrm{II}} &= (2.7 \pm 0.1) \cdot 10^{-7}.
\end{align*}
The PSD of the type-II MOT exceeds that of the type-I MOT by more than two orders of magnitude.\\
The setup also provides an atom source and laser system for cooling and manipulation of $^{41}$K and $^{39}$K which have most spectroscopic features of $^{87}$Rb in common\cite{Tiecke}. In contrast to $^{87}$Rb, they exhibit narrow hyperfine splittings of D2-transitions, which complicates laser cooling of $^{41}$K and $^{39}$K in standard type-I MOTs.\\
By adapting the above described changes in the potassium laser system, the feasibility of a blue-detuned type-II MOT for cooling of $^{41}$K and $^{39}$K is explored. 
For both isotopes, cooling and trapping are only achieved if solely the polarization of the blue-detuned repumping light field is reversed. The cooling laser remains in the red-detuned type-I configuration. 
The experiments show neither higher densities nor lower temperatures for the mixed type-I/II MOT, indicating no advantages compared to the pure type-I MOT. This is attributed to additional heating due to simultaneously driven type-I transitions within the narrow level splitting of potassium. 
\section{Conclusion}
The setup and measurements presented in this paper demonstrate the feasibility of a blue-detuned type-II MOT by utilizing LCVRs for fast polarization control. This allows for a low-cost implementation in existing experimental setups without the need of additional lasers or beam collimators. \\
By switching from red-detuned type-I to blue-detuned type-II transitions with reversed polarizations, 85\% of the $^{87}$Rb atoms trapped in the type-I MOT were transferred into the type-II MOT. TOF measurements show that a geometric mean temperature of $\overline{T}_\mathrm{II}=44.5\,$\textmu K of the trapped  was reached. This is well below the Doppler temperature $T_D\approx 140\,$\textmu K which typically constitutes the lower temperature limit within a type-I MOT. Furthermore, the PSD of the type-II MOT is shown to be increased by more than two orders of magnitude compared to the type-I MOT. Among other methods like dark spontaneous-force MOTs \cite{Ketterle1993, Radwell2013}, blue-detuned type-II MOTs offer an alternative and easy-to-implement route to effectively increase the PSD of trapped ensembles. \\
A type-II MOT combines the advantages of a sub-Doppler cooled optical molasses with the convenient position control of a trapped ensemble. Hence, type-II MOTs might also help to improve the mode matching between the MOT and consecutive stages of trapping and cooling, in particular for magnetic or optical dipole traps. Accordingly, center-of-mass oscillations of transferred ensembles due to positional mismatch could be reduced to a minimum compared to a free-falling optical molasses.\\
LCVRs prove to be a simple and compact device to efficiently improve the atom cooling performance. This applies especially for setups with rigorous constraints on space, power consumption and weight like experiments on mobile \cite{Hinton2017,Bidel2018} or microgravity platforms \cite{Elsen2021, Rudolph2015, Frye2021}.\\

\begin{acknowledgments}
We would like to thank Thijs Wendrich and Wolfgang Bartosch for electronic support. We are grateful to Maike Lachmann for her experimental support and comments on this manuscript.\\
This work is supported by the DLR Space Administration with funds provided by the Federal Ministry for Economic Affairs and Energy (BMWi) under grant number DLR 50WP1431, and is funded by the Deutsche Forschungsgemeinschaft (DFG, German Research Foundation) under Germany’s Excellence Strategy – EXC-2123 QuantumFrontiers - 390837967. 
\end{acknowledgments}
\section*{Data availability}
The data that support the findings of this study are available from the corresponding author upon reasonable request.

\nocite{*}
\bibliography{aipsamp}

\end{document}